\documentclass{iopart}
\usepackage{graphicx}
\begin{document}

\letter{Generating ring currents, solitons,
and svortices by stirring a Bose-Einstein condensate in a toroidal trap} 
\author{Joachim Brand and William P. Reinhardt}
\address{Department of Chemistry, University of Washington, Seattle,
WA 98195-1700, USA} 
\date{\today}
\eads{\mailto{joachim@ringb.chem.washington.edu},
\mailto{rein@chem.washington.edu}} 
\submitto{\JPB}

\maketitle

\begin{abstract}
We propose a simple stirring experiment to generate quantized
ring currents and solitary excitations in Bose-Einstein condensates in
a toroidal trap geometry. Simulations of the 3D Gross-Pitaevskii
equation show that pure ring current states can be generated
efficiently by adiabatic manipulation of the condensate, which can be
realized on experimental time scales. This is illustrated by
simulated generation of a ring current with winding number two. 
While solitons can be generated
in quasi-1D tori, we show the even more robust generation of hybrid,
solitonic vortices (svortices) in a regime of wider
confinement.  Svortices are vortices confined to essentially
one-dimensional dynamics, which obey a similar phase-offset--velocity
relationship as solitons. 
Marking the transition between solitons and vortices, svortices are a
distinct class of symmetry-breaking stationary and uniformly rotating
excited solutions of the 2D and 3D Gross-Pitaevskii equation in a
toroidal trapping potential. Svortices should be observable in dilute-gas
experiments.
\end{abstract}
\nosections
Persistent currents, quantized vortices, and solitons in Bose-Einstein
condensates (BEC) are high-energy excitations predicted by mean-field
theory. Quite recently, both solitonic waves
\cite{denschlag00,burger99} and singly quantized vortices
\cite{matthews99,anderson00,madison00} have been unambiguously
identified experimentally in 3D harmonic trap geometries. The next
step in increasing our understanding of quantized currents and
solitons is to study their stability properties and measure their
intrinsic lifetimes thereby testing the validity of mean-field theory
in a so-far unexplored regime.  The torus presents an ideal geometry
in this context as solitons can travel without reaching an end of the
trap and ring currents are stabilized by the ring geometry as opposed
to vortices where the core can drift \cite{drift99}.  Excited states
in toroidal traps also have the potential to provide useful
applications. E.g.~Drummond {\it et.~al.}
\cite{drummond00ep} have suggested constructing a mode-locked atom
laser using a dark soliton with a nodal plane (also called a black
soliton or kink configuration) in a ring geometry.  They demonstrated
the creation of a black soliton by stirring a condensate in a
1D ring with a paddle at constant angular velocity and
damping additionally generated excitations by Raman outcoupling.
Other simulations involving nonlinear excitations in toroidal
geometries can be found in references \cite{feder00pp,martikainen00ep}.

In this letter we propose an adiabatic stirring technique to impose
circulation onto a BEC in a 3D toroidal geometry. Without any need for
external cooling or damping mechanisms, adiabatic manipulations allow
one to generate pure ring-current states with essentially arbitrary
circulation. A variation of the stirring technique can produce
solitons if the transverse trap confinement in the toroidal tube is
quasi-1D\footnote{We define a trapping geometry quasi-1D when the
confinement in two of three spatial dimensions is such that the 1D
Gross-Pitaevskii equation essentially governs the dynamics, which is
the case for transverse confinements of the order of the healing
length \cite{carr00:quasi1D,petrov00}.}.  For slightly less tight
confinement new nonlinear objects are created, which have properties
of both vortices and solitons. We call these objects solitonic
vortices, or svortices, and briefly discuss their properties.

In our simulations, the condensate is stirred with a rotating paddle
which can easily be realized with a blue-detuned laser-generated light
sheet. Being purely optical in nature, this
method does not rely on specific magnetic field configurations and is
suitable for single-component as well as spinor
condensates. Toroidal trap
geometries have already been used in early BEC experiments
\cite{davis95} and in the recent JILA vortex experiments
\cite{matthews99,anderson00}.  Quasi-1D confinement in a toroidal trap
certainly presents a challenge for experimentalists but seems
realistic after promising results in linear geometries \cite{bongs00}
and with the prospect of a tightly confining purely optical toroidal
trap \cite{wright00}.

We model the dynamics of the BEC with the time-dependent
Gross-Pitaevskii or nonlinear Schr\"odinger equation (NLSE), which
provides the zero-temperature mean-field theory
\cite{dalfovo99}. For a BEC of $N$ atoms of mass $M$, confined in an
external potential $V$, the NLSE reads:
\begin{equation} \label{nlse}
  i\hbar \partial_t \psi = [-(\hbar^2/2M)\nabla^2 + V +
  g|\psi|^2]\psi ,
\end{equation}
where $g = 4\pi \hbar^2 a_0 N/M$ is the coupling constant. The
$s$-wave scattering length $a_0$ is assumed to be positive, relating
to repulsive interparticle interactions. The order parameter
or condensate wavefunction $\psi$ is normalized to $\int |\psi|^2\,
$d{\bf r} $= 1$ and may be written as
%\begin{equation}
$ \psi(${\bf r}$,t) = \sqrt{\rho}\, \exp{i\phi}$,
%\end{equation}
where the square root of the density $\sqrt{\rho}$ and the phase
$\phi$ are real functions of {\bf r} and $t$.

We have solved the time-dependent NLSE (\ref{nlse}) numerically on a
grid using a pseudo-spectral FFT representation and a 4th-order
variable-step Runge-Kutta integrator in box boundary conditions.  The
trap potential $V = V_{\rm ho} + V_{\rm co} + V_{\rm pa}(t)$ mimics a
torus by a 3D harmonic-oscillator potential $V_{\rm ho}$ in the shape
of an oblate disk pierced by a Gaussian blue-detuned laser beam
modelled by the core potential
$V_{\rm co}$ resembling the trap used in early MIT experiments
\cite{davis95}. A time-dependent ``paddle'' potential $V_{\rm pa}(t)$
mimicking an ellipsoidal Gaussian light sheet is
used to stir the condensate in the torus\footnote{In detail, the
harmonic part of the potential is $V_{\rm ho}
= M (\omega_x x^2 + \omega_y y^2 + \omega_z z^2)/2$ with $\omega_x =
\omega_y = 0.4\,\omega_z$; the core-potential is $V_{\rm co} =
V_{\rm c} \exp\{-(x^2 + y^2)/(2\Delta r_{\rm c}^2)\}$. The paddle potential
has the form of a Gaussian ellipsoid centred at a radius $r_{\rm p}$
and initially reads $V_{\rm pa}(0) = V_{\rm p} \exp\{-(x-r_{\rm
p})^2/(2\Delta r_{\rm p}^2) - y^2/(2\Delta a_{\rm p}^2)\}
\theta(x)$. During the simulation it is accelerated to a uniform
rotation around the $z$-axis and retracted by pulling out to the side.
The following parameters $\omega_x = 
                  2\pi\, 8.63$Hz, $V_{\rm c} = 18.5$nK, $\Delta r_{\rm
                  c} = 15.1\mu$m, 
                  $V_{\rm p} = 6.95$nK, $\Delta a_{\rm p} =
                  3.37\mu$m, $\Delta r_{\rm p} =
                  27.0\mu$m, $r_{\rm p} = 32.4 \mu$m, were used in the
simulation and are comparable to currently achievable experiments.
}.

\begin{figure}
\begin{center}
\includegraphics{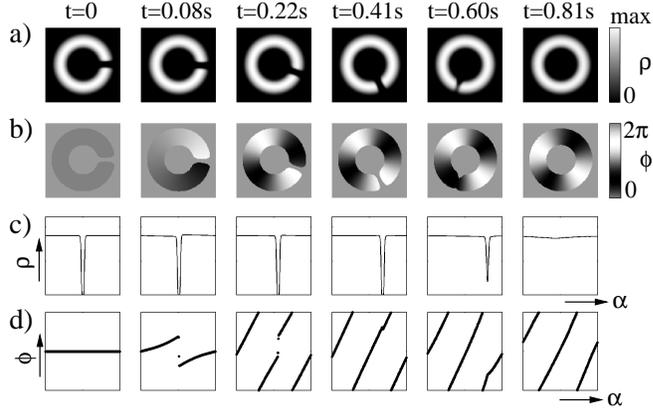}
\end{center}
\caption{3D simulation of ring current generation with $w = 2$. Parts a)
and b) show gray-scale coded 2D slices of the density $\rho$ and the phase
$\phi$ respectively in the $xy$ plane of the trap. Rows c) and d) show
the density and the phase on the line of minimal potential
parameterized by the angle $\alpha$. Note that the phase is plotted
within an arbitrary offset and modulo $2\pi$. The columns show time
slices of our simulation. The first column shows the condensate
ground state in the toroidal trap intersected by a resting stirring
paddle.  Columns 2 and 3 show how a phase gradient builds up while the
stirring paddle is accelerated at a constant rate between $t=0$ and
$t=0.27$s. Afterwards the condensate is stirred with constant angular
velocity $\omega = 2\pi\, 0.10$Hz. At this stage, the density notch
goes down to about 1\% of the maximum density and a phase step of
$\Delta \phi \approx 4\pi$ has built up across the paddle.  The fifth
column shows how the paddle is retracted between $t=0.41$s and
$t=0.68$s. The last column shows the $w=2$ ring current that has
established in the toroidal trap. The transverse confinement in the
torus is $\approx 25$ healing lengths.}
\label{fig_b48_vort_gen}
\end{figure}

For a condensate of $10^6$ atoms of Na with $a_0 = 2.75$nm, our
simulations relate to a torus of radius $r_{\rm T} = 32.4 \mu$m
(distance of the potential minimum from the symmetry axis of the
torus).  The healing length $\xi(${\bf r}$) = 1/\sqrt{8\pi a_0 N
\rho(\mbox{\bf r})}$, which depends on the spatially inhomogeneous
density $\rho(\mbox{\bf r})$, sets the relevant length scales for the
size of solitons ($\approx 2\xi$) and the transverse size of vortex
filaments ($\approx 3\xi$) \cite{kivshar98}. From the numerically
obtained ground state wavefunction in the torus, we find a value of $\xi
\approx 2.2\mu$m for the size of the healing length at peak
density. The transverse confinement of the condensate in the toroidal
tube is best characterized by the number of healing lengths, which we
define by the line integral ${\cal N}_{\xi} =
\int_{{\cal C}}\xi(\mbox{\bf r})^{-1}\, {\rm ds}$, taken across the
toroidal tube along a radial line ${\cal C}$ embedded in the symmetry plane of
the torus.  In other geometries, the most sensible curve of
integration ${\cal C}$ is dictated by the confinement. 
The dimensionless confinement parameter ${\cal
N}_{\xi}$ can be computed directly form the numerical wavefunction and
does not rely on the Thomas-Fermi approximation, which is often used
to characterize the extent of a trapped condensate. In our simulation the
confinement in the radial direction is about the same as in the axial
direction and amounts to ${\cal N}_{\xi} \approx 25$.
Figure~\ref{fig_b48_vort_gen} shows the result of the simulation of a stirring
experiment to generate a ring current with winding number $w=2$.  The
ground state of the condensate in the toroidal trap
intersected by a narrow ellipsoidal Gaussian paddle potential
was found by imaginary-time
propagation.  During the simulation the paddle was accelerated to
azimuthal rotation ``stirring'' the condensate and subsequently
retracted in the direction of the torus axis.

In order to elucidate the effect of adiabatic stirring on the
condensate wavefunction and to interpret the result of the simulation
shown in figure~\ref{fig_b48_vort_gen}, 
we start with an idealized 1-D picture, neglecting the
transverse dimensions of the torus: When blocked by
a paddle, the torus can be treated as a 1D box with hard-wall boundary
conditions. The ground state $\psi_g$ of the condensate in a 1D box is
a Jacobi elliptic function of type $\mbox{sn}$ \cite{carr00:1}, which
is real-valued and therefore has a constant phase. When the paddle is
moving, the wavefunction acquires a phase ramp. This can be understood
by considering the Galilean invariance of the NLSE: The boosted ground
state $\psi_g(x-vt,t)\exp[iM/\hbar(vx - \frac{1}{2} v^2
t)]$ satisfies the boundary conditions in the moving frame and is the
``ground state'' of the translated box. A phase gradient in the
wavefunction is equivalent to a supercurrent with velocity $v = \hbar
\partial_x \phi / M$.  A state like this can either be reached by an
adiabatic transition between the resting and the moving box through
accelerating the paddle slowly or one can prepare the condensate right
from the start in a torus with a rotating paddle.  Removal of the
stirring paddle reconnects both ends of the ``1D box'' to a ring and
can be modelled in 1D as decreasing a narrow potential barrier to zero
height. In the adiabatic limit of this process, the phase step of
$\Delta \phi = 2\pi r_{\rm T} v M/\hbar^2$ acquired by stirring
reconnects at the nearest integer multiple of $2\pi$ while the density
is still low. In this way phase gradients are smoothed out before the
density notch fills in completely leaving the condensate in an
azimuthally symmetric current-carrying state thus creating a pure
toroidal ring current. In this idealized picture there is no upper
limit for the current velocities. Above a critical velocity, however,
the supercurrent is instable against phase slips on resting obstacles
which might arise through imperfections in the trap \cite{hakim97}.

It is not obvious that the simple 1D picture of adiabatic stirring
sketched above should be transferable to experimental conditions in 3D
tori well out of the quasi-1D limit.  The simulation of
figure~\ref{fig_b48_vort_gen}, however, shows that an essentially pure
ring-current state can be generated by realistic choices of time
scales and parameters.
In order to get a rough estimate for the timescale of adiabaticity,
consider the time $t_{{\rm ad}}= 2\pi\, r_{{\rm T}}/c_{\rm max}$ it
takes for a sound wave to travel around the ring of radius $r_{{\rm
T}}$ once. We use the Bogoliubov speed of sound $c_{\rm max} =
\sqrt{4\pi\hbar^2 a N \rho/M^2}$ \cite{bogoliubov47}, which amounts to
$c_{\rm max} = 0.89 {\rm mm\,s}^{-1}$ at the peak density of the
numerically obtained ground-state wavefunction, to estimate this time
to $t_{{\rm ad}} \approx 0.09$s. It is further useful to compare the
acceleration rate of the stirring paddle to $a_{\rm ad} = c_{\rm
max}^2/(2 \pi r)$, which is the acceleration that brings the
velocity up to the speed of sound in the time that it takes a sound
wave to travel around the ring. We find that acceleration rates
of about $0.1\, a_{\rm ad}$ work well. Alternatively,
instead of starting with the ground state of an intersected toroidal
trap, 
the light sheet potential barrier can also be ramped up slowly. Our
simulations show that ramping up the paddle linearly within  about $3
t_{{\rm ad}}$ is sufficiently slow in order to make no appreciable
difference for the generation of ring currents. Further,
retracting the core potential $V_{\rm co}$, as
realized in reference~\cite{anderson00}, leads
to the generation of vortex filaments. We observe
that essentially pure vortex states can be generated by ramping down the
core-potential on a timescale of about $2 t_{{\rm ad}}$.

A variation of the stirring experiment as described above can be used
to generate solitons.
In the non-adiabatic limit, when the intersecting light sheet is
removed suddenly, the imposed phase step and density notch set an
initial condition for the dynamics of the condensate. It has been
discussed in other places, that the combination of a phase step
\cite{denschlag00} and a density notch is a very efficient way to
start density-notch solitons \cite{reinhardt97,carr00pp}.  For the
example of a linear quasi-1D trap it is discussed by Carr {\it
et.~al.} \cite{carr00pp} how the combination of a density notch
generated by sudden removal of a Gaussian potential barrier in
combination with a phase step can lead to the generation of one or
more solitons with different velocities depending on the size of the
phase step and the width of the potential barrier. 
We have performed simulations in tightly-confined toroidal traps with
a transverse confinement in terms of healing lengths ${\cal N}_{\xi}$
between 2 and 5, where we observe a quasi-1D type of behaviour and the
discussion in reference~\cite{carr00pp} applies. Our 3D simulations show the
generation of band solitons upon sudden removal of the paddle.  

\begin{figure}
\begin{center}
\includegraphics{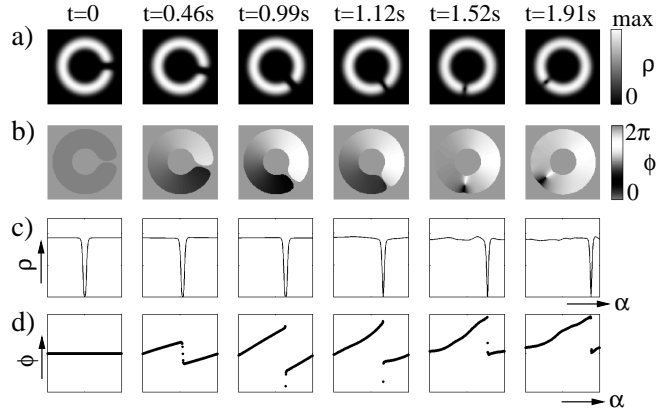}
\end{center}
\caption{Stirring experiment where a svortex excitation is
generated. Shown are the density $\rho$ and the phase $\phi$ of 
the condensate at different times as in
figure~\ref{fig_b48_vort_gen}. The first and the second column show
the acceleration phase which lasts from $t= 0$ to $t =
0.66s$. Afterwards the condensate is stirred with constant angular
velocity $\omega = 2\pi\, 0.032$Hz. The paddle is contracted between
$t=0.99s$ and $t=1.15s$. Due to the transverse confinement being about
9 healing lengths, a dark soliton is not completely stabilized but
transforms into a vortex by the influence of curvature. A state is
generated which is on the transition between a dark soliton and
vortex.}
\label{fig_sol_gen}
\end{figure}

\begin{figure}
\begin{center}
\includegraphics{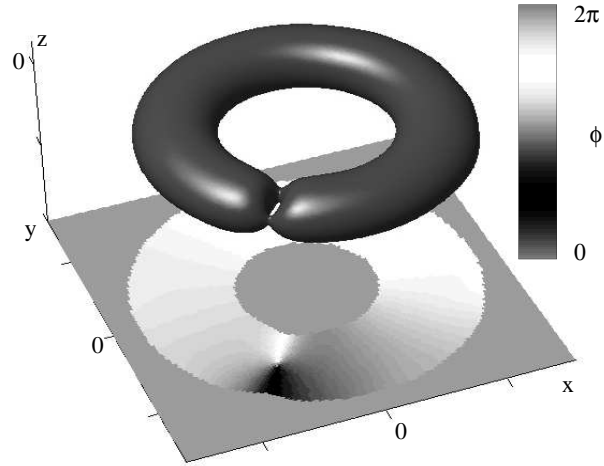}
\end{center}
\caption{An unusual type of solitary excitations, which we call
a svortex, can be generated in toroidal BECs by stirring. Shown are the
iso-surface of the condensate at 16\% of the maximum density and the
phase $\phi$ in the xy plane. The condensate wavefunction was prepared in a
stirring experiment described in the text below. The solitary density
deformation and phase singularity preserve their shape while rotating
clockwise about the axis of the torus with roughly half the angular
velocity of a ring current carrying a single quantum of vorticity. The
current plot shows the wavefunction at $t = 1.98$s. }
\label{fig_svort_iso}
\end{figure}

However, in
situations where the confinement is less tight and the curvature of
the torus becomes apparent the phenomenology changes:
We observe the robust generation of a unique class of solitary
excitations in a torus with transverse confinement in
terms of healing lengths ${\cal N}_{\xi} \approx 9$.
These excitations have properties of both solitons and vortices,
as will be elaborated below, and therefore
we call them solitonic vortices or {\it svortices}.
Figure~\ref{fig_sol_gen} shows a simulated stirring experiment
in a condensate of $10^6$ atoms of Na
where a phase step of
approximately $\pi$ is generated by stirring\footnote{We use the same
potential as before with the following parameters for a condensate of
$10^5$ Na atoms: $\omega_x = 
2\pi\,21.0$Hz, $V_{\rm c} = 135$nK, $\Delta r_{\rm c} = 16.8\mu$m;
$V_{\rm p} = 50.6$nK; $\Delta a_{\rm p} = 3.75\mu$m; $\Delta r_{\rm p}
= 30.0\mu$m. The toroid radius in this simulation is $r_{\rm T} =
36.0\mu$m. At peak density of the condensate ground state in the
torus we find a Bogoliubov sound
speed of $c_{\rm max} = 2.4 {\rm mm\,s}^{-1}$ and
healing length of $\xi = 0.82\mu$m.
}. Ramping down the paddle
potential within 66ms (which is roughly $1/3\, t_{\rm ad}$ with
$t_{\rm ad} \approx 0.23$s in this simulation), a solitary wave is
created that has characteristics of both solitons and vortices. 
While showing the  phase singularity signature of a vortex, the
density profile is squeezed by the tranverse confinement (see
figure~\ref{fig_svort_iso}). After the stirring potential is turned off,
the svortex moves at constant angular velocity around the ring. Our
simulations show that the velocity of the svortex
is sensitive to the details of the stirring process like
the paddle acceleration rate and timescale of paddle retraction.
Similar to solitons, the velocity of svortices relative to the
condensate is related to a phase offset.

\begin{figure}
\begin{center}
\includegraphics{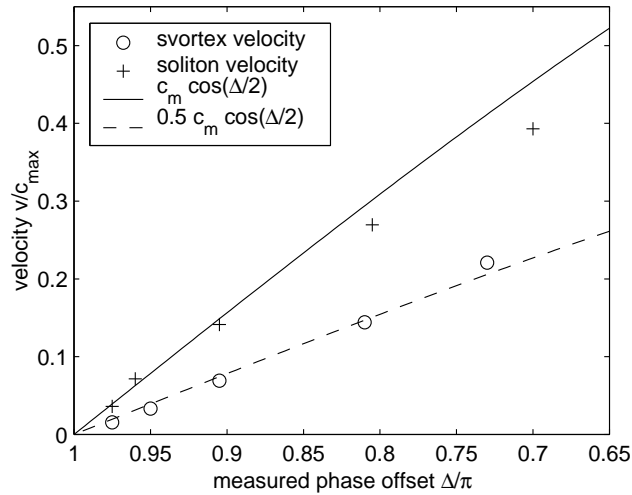}
\end{center}
\caption{Velocity vs.~phase-offset relation for a band soliton and a
svortex in a 2D box with dimensions $8\xi \times 64\xi$. A stationary
black band soliton (nodal-plane state) and a stationary svortex were
set in motion by imprinting a phase step. The deviation of the
measured phase offset $\Delta$ from $\pi$ is approximately half the
imprinted phase step.}
\label{fig_velo}
\end{figure}

The velocity of solitons relative to the condensate is given by 
\begin{equation} \label{velooffset}
v = c_{{\rm max}} \cos(\Delta/2)
\end{equation}
\cite{denschlag00,kivshar98,reinhardt97}, where $\Delta$ is the phase
step across the density notch relating to standing black solitons
(kink states) at $\Delta = \pi$ and to increasingly shallow gray
solitons moving with the speed of sound $c_{{\rm max}}$ as $\Delta$
approaches zero. The experiments of reference~\cite{denschlag00} have
confirmed this relation for dark solitons in a BEC. We find that
svortices obey a similar relationship. Figure~\ref{fig_velo} shows the
velocity--phase-offset relation for a svortex and a band soliton,
which have been created by imaginary-time propagation in a long thin
rectangular 2D box (of dimensions $8\xi \times 64\xi$) with hard-wall
boundary conditions \cite{brand00tbp}. We find that the svortex
velocity depends linearly on small deviations of the phase offset from
$\pi$ resembling the relation (\ref{velooffset}) for solitons.
The proportionality coefficient, however, is below the value for
solitons, given by the Bogoliubov sound speed $c_{{\rm
max}}$. Simulations in differently sized boxes show that this
coefficient depends strongly on the transverse confinement. We observe
the discussed behaviour for transverse confinements ${\cal N}_{\xi}$
between 6 and 12 healing lengths. It is this velocity--phase-offset
relation, which makes the vortex confined to a thin waveguide a
solitonic vortex or svortex. A detailed analysis of the
svortex's properties including the collisional properties, which are
considerably more complicated than those of solitons, will be
published elsewhere
\cite{brand00tbp}.  We note that the term "vortex soliton" or "optical
vortex soliton" has been used before in the nonlinear optics community
in a much broader context \cite{kivshar98} whereas we introduce the
term solitonic vortices referring to transversely confined vortices
with specific solitonic properties.

In conclusion, we have shown that ring currents and solitonic
structures can be generated very efficiently in toroidal trap
geometries by stirring the condensate with a blue detuned laser light
sheet and consequently retracting the stirring paddle. Pure
ring currents can be generated this way for a wide range of transverse
confinement parameters. Density notch dark and gray solitons can be
generated in quasi-1D transversely confined tori.  In a regime of
slightly wider confinement, a novel hybrid object with properties of
solitons and vortices is generated.  These solitonic vortices show a
phase-offset--velocity relationship similar to solitons.

%\ack
We like to thank L.~Carr and S.~McKinney at the University of
Washington and D.~Feder and J.~Denschlag at NIST for helpful
discussions and comments. This work was partially supported by NSF
Chemistry and Physics.  J.~B.~acknowledges support from the
Alexander von Humboldt Foundation through a Feodor Lynen Research
Fellowship.

\section*{References}

%\bibliographystyle{prsty}
%\bibliography{refs,refs_bec,notes}

\end{document}